# Imaging the Ettingshausen effect and cryogenic thermoelectric cooling in a van der Waals semimetal

T. Völkl[1][†], A. Aharon-Steinberg[1][†], T. Holder[1,2], E. Alpern[1], N. Banu[1], A. K. Pariari[1], Y. Myasoedov[1], M. E. Huber[3], M. Hücker[1], and E. Zeldov[1]*

[1]Department of Condensed Matter Physics, Weizmann Institute of Science, Rehovot 7610001, Israel

[2]School of Physics and Astronomy, Tel Aviv University, Tel Aviv 69978, Israel

[3]Departments of Physics and Electrical Engineering, University of Colorado Denver, Denver, Colorado 80217, USA

[†]These authors contributed equally to this work

*eli.zeldov@weizmann.ac.il

Attaining viable thermoelectric cooling at cryogenic temperatures is of major fundamental and technological interest for novel electronics and quantum materials applications. In-device temperature control can provide a more efficient and precise thermal environment management as compared to the conventional global cooling. Here we develop nanoscale cryogenic imaging of a magneto-thermoelectric effect and demonstrate absolute cooling and an ultrahigh Ettingshausen effect in exfoliated $WTe_2$ Weyl semimetal flakes at liquid He temperatures. Application of a current and perpendicular magnetic field gives rise to cooling via generation of electron-hole pairs on one side of the sample and heating by their recombination at the opposite side. In contrast to bulk materials, the cooling process is found to be nonmonotonic in magnetic field and in device size. The derived model of magneto-thermoelectricity in mesoscopic semimetal devices shows that the cooling efficiency and the induced temperature profiles are governed by the interplay between sample geometry, electron-hole recombination length, magnetic field, and flake and substrate heat conductivities. The findings open the way for direct integration of microscopic thermoelectric cooling and for temperature landscape engineering in novel van der Waals devices.

Thermoelectric effects have long been used for power generation, cooling of electronic devices, and temperature measurements, where compact size, high stability, vibrationless operation without moving parts, and high tunability are required [1,2]. Historically, research efforts have focused on elevated temperatures since the cooling efficiency decreases significantly towards low temperature, such that it is eventually overshadowed by Joule heating for realistic current strengths. In recent years, the advent of high-purity single crystal semimetals has led to increased interest in thermoelectric effects at cryogenic temperatures [3–12].

The common approach to thermoelectric cooling is based on the Peltier effect [13], in which a temperature gradient is attained along the current flow direction in structures composed of regions or materials with different Peltier coefficients, thus carrying different average amount of heat per transferred unit charge. Peltier cooling in bulk materials down to liquid nitrogen temperatures has been realized in a number of studies [2,14], with only one report of bulk cooling at liquid He temperatures [15]. By engineering microscopic devices that form energy gaps in the electron density of states, like superconducting junctions or quantum dots, the efficiency of Peltier cooling at very low temperatures has been shown to enhance significantly due to preferential transmission of electrons at specific energies [16–18].

Here we address a very different and interesting magneto-electro-thermal mechanism, the Ettingshausen effect [19], which has not been studied in microscopic devices previously [18], and provide the first observation of the Ettingshausen effect and of Ettingshausen cooling at 4 K. The application of a charge current in presence of perpendicular magnetic field leads to a temperature gradient that is transverse to the current flow direction, rather than along it. In a metal, this effect is due to the energy dependence of the drift velocity of the charge carriers, which in the presence of magnetic field, deflects the high- and low-energy carriers to opposite sides of a sample, thus providing heating and cooling upon equilibration of the carriers with the lattice (Fig. 1a). The resulting current to temperature gradient conversion in this case is rather weak.

A much stronger effect can be attained in semiconductors and semimetals hosting both electrons and holes with approximately equal densities and mobilities [18,20]. Under such circumstances, the Hall voltage that usually develops across a device is diminished and the unbalanced Lorentz force drives both the electrons and holes in the same transverse direction (Fig. 1b). This leads to a nonequilibrium electron-hole pair accumulation on one side and depletion on the opposite side. As a result, recombination and generation of electron-hole pairs is increased, thus providing heating and cooling at opposing sides through equilibration with the lattice [19]. As such, heating and cooling is expected to be constrained to a small area within a distance of the recombination length $l_R$ from the boundaries. Hence, a thermal measurement technique with spatial resolution on that length scale is required in order to fully elucidate understanding of the effect.

The search for suitable materials for thermoelectric cooling has focused predominantly on bulk systems, leaving the unrecognized potential of mesoscopic devices for integrated cooling largely unexplored [18]. Employing van der Waals (vdW) semimetals as thermoelectric elements [21–26], is especially promising since they can be readily integrated into stacks of atomic layers and moiré heterostructures [27,28]. Herein, different materials such as insulating, semiconducting, superconducting, or magnetic layers in the heterostructure can fulfill different functionalities. Accordingly, we propose the use of a vdW semimetal as an active thermoelectric element for in-device cooling or for engineering a controllable temperature gradient within a device. A particularly promising vdW material for thermoelectric cooling is the Weyl semimetal $WTe_2$, because of its very high charge carrier mobility and near compensation of electron and hole densities [29–31]. In its bulk form, it has already been shown to display a large Nernst effect [32] and ultrahigh Ettingshausen effect at temperatures above 20 K [12].

Thermoelectric mechanisms in which out-of-equilibrium voltages or currents are generated by temperature gradients, like the Seebeck and Nernst effects, have been investigated extensively in microscopic vdW devices. These studies are typically performed by measurement of voltages across microscopic contacts induced by global temperature gradients [21–26,33–35], or by local heating using e.g. a scanning focused laser beam [36–45]. In contrast, electro-thermal processes, like the Peltier and Ettingshausen effects, for which temperature gradients need to be measured or imaged in response to applied currents, are much more challenging to investigate in microscopic devices, and have thus far been restricted to above liquid nitrogen temperatures [46,47]. In this work, we provide thermal imaging of magneto-electro-thermal cooling at liquid He temperatures in exfoliated flakes of the transition metal dichalcogenide $WTe_2$, revealing ultrahigh Ettingshausen effect and the underlying mesoscopic mechanisms that have been inaccessible experimentally hitherto.

**Cryogenic thermal imaging technique**

A superconducting quantum interference device on a tip (SQUID-on-tip, SOT) [48] made of MoRe [49] with diameter of 110 nm was scanned above the sample surface at a height $h \cong 80$ nm in He exchange gas atmosphere. It was employed as a nanothermometer [50] with 2.6 µK/Hz$^{1/2}$ sensitivity at $T_0 = 4.3$ K, operating in an applied out-of-plane magnetic field of up to $B = 5$ T (Methods), as shown schematically in Fig. 1b. A high-quality $WTe_2$ single crystal [31] was exfoliated to produce a $d = 260$ nm thick flake and patterned by reactive ion etching (Methods) into a number of rectangular chambers with different widths $W$ (Fig. 1c). To circumvent the $1/f$ noise of the SOT, a sine or square-wave *ac* current at frequency $f = 85.37$ Hz and variable rms amplitude $I$ was applied to three chambers connected in series by narrow constrictions (Fig. 1c), and the resulting *ac* change in the local temperature $\delta T(x,y)$ relative to the base temperature $T_0$ was imaged by the scanning SOT.

**Imaging the Ettingshausen effect in $WTe_2$**

The current-induced change in the local temperature $\delta T$ has two contributions, one arising from Joule heating, $\delta T_J$, and the other from the Ettingshausen effect, $\delta T_E$. By applying a sine-wave *ac* current through the sample, we can image these two contributions independently as follows. Since the Ettingshausen effect is linear in $I$, $\delta T_E$ is a sine-wave temperature modulation at frequency $f$ of the applied current. We thus detect this contribution by lock-in measurement at the fundamental frequency $f$. In contrast, Joule heating is quadratic in $I$ and therefore results in *dc* heating superimposed by an *ac* temperature modulation at frequency $2f$ (Methods). We image this *ac* $\delta T_J$ modulation by lock-in detection at the second harmonic.

Figures 1d-f show the Joule-heating-induced $\delta T_J(x,y)$ in the largest chamber (dashed rectangle in Fig. 1c) at $B = 0$, 1.04, and 5 T respectively at $I = 50$ µA. $\delta T_J$ is rather uniform in most of the chamber area with intense hot spots at the constrictions where the current density is high. Figure 1j shows that $\delta T_J$ in the chamber grows quadratically with $I$, as expected for Joule heating. $\delta T_J$ increases by over two orders of magnitude with field, growing approximately quadratically with $B$, as shown in Fig. 1k. This is the consequence of a very large magnetoresistance, $\rho_{xx}(B) \propto B^2$, in high-purity $WTe_2$ as reported previously [29–32,51].

The Ettingshausen temperature distribution $\delta T_E(x,y)$ is shown in Figs. 1g-i. At zero field, $\delta T_E = 0$ as expected (the weak signal in the constrictions is an artifact, see Methods). Upon increasing the field, a large temperature gradient develops transverse to the current and the magnetic field directions. The temperature difference, $\Delta T_E = \delta T_E\left(0, -\frac{W}{2}\right) - \delta T_E\left(0, \frac{W}{2}\right)$, between the hot (bottom) and the cold (top) edges at $y = \mp\frac{W}{2}$ across the sample center ($x = 0$), grows linearly with $I$ (Fig. 1j) as expected for the Ettingshausen effect, $\Delta T_E = P_E IB/W$, where $P_E$ is the Ettingshausen coefficient. For $B = 5$ T and $W = 20$ µm, we find an ultrahigh Ettingshausen signal $\Delta T_E/(JW) = 2.2\cdot10^{-5}$ KA$^{-1}$m ($J$ is the current density) comparable to the previously reported bulk values [12], but at a much lower temperature. However, in contrast to the standard bulk behavior, we find that $\Delta T_E$ is not linear in $B$ as shown in Fig. 1l. Even more surprisingly, investigating chambers with different widths reveals a nonmonotonic $W$ dependence, with the chamber with $W = 10$ μm showing a larger $\Delta T_E$ than those with $W = 20$ μm and $W = 5$ μm (Fig. 1l). Moreover, Figs. 1h,i show a nontrivial spatial distribution of $\delta T_E(x, y)$ with enhanced heating and cooling signals on the left and right edges near the constrictions. These features are also observed in additional samples as shown in Extended Data Figures 2 to 4.

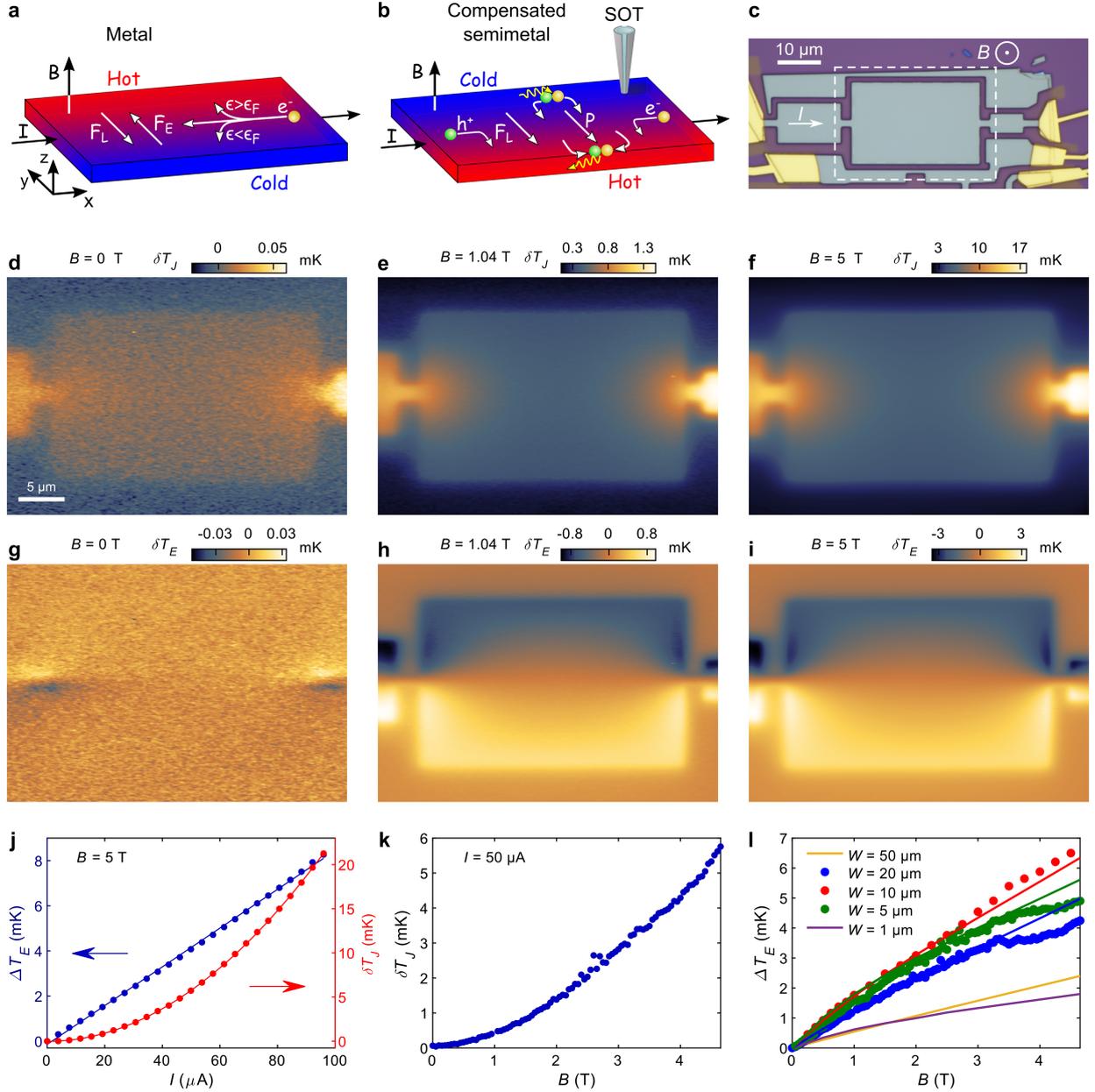

**Fig. 1. Thermal imaging of the Ettingshausen effect in WTe₂ flake at 4.3 K.** (**a,b**) Schematics of the Ettingshausen effect in metals (a) and in compensated semiconductors or semimetals (b) in the presence of perpendicular magnetic field $B$. In a metal, the average Lorentz force $\boldsymbol{F}_L$ is counterbalanced by the transverse electrostatic force $\boldsymbol{F}_E$ of the Hall voltage. Since the higher energy electrons usually have a lower drift velocity and hence lower $\boldsymbol{F}_L$, a small transverse thermoelectric effect develops. In semiconductors and semimetals, a much larger transverse temperature gradient of opposite sign is induced by electron-hole pair generation at the cold edge and recombination at the hot edge. (**c**) Optical image of WTe₂ device showing three rectangular chambers (grey) of width of $W = 10, 20$, and 5 μm along with surrounding additional patterns. A sinusoidal $ac$ current $I$ at frequency $f$ is applied in series (white arrow) through the Au contacts (yellow). The dashed white rectangle marks the scan window in (d-i). (**d-f**) Temperature maps $\delta T_J(x, y)$ corresponding to Joule heating acquired at $2f$ with $I = 50$ μA at magnetic fields of $B = 0$, 1.04, and 5 T. (**g-i**) Corresponding maps of Ettingshausen temperature component $\delta T_E(x, y)$ acquired simultaneously at $f$. (**j**) Left axis: temperature difference between the hot and cold edges $\Delta T_E$ (blue dots)

vs. $I$ with a linear fit (blue line) at $B = 5$ T. Right axis: $\delta T_J(0,0)$ at the center of the chamber vs. $I$ (red dots) and a parabolic fit (red line). (**k**) Magnetic field dependence of $\delta T_J$ at $I = 50$ μA. (**l**) $\Delta T_E$ vs. $B$ in the three adjacent chambers of $W = 5$, 10, and 20 μm at $I = 50$ μA (dots) showing nonlinear dependence on $B$ and nonmonotonic dependence on $W$. The solid curves are the corresponding $\Delta T_E$ from the numerical simulations.

**Absolute cooling**

The linear increase of the Ettingshausen temperature difference $\Delta T_E$ with current does not necessarily mean that the temperature of the cold edge of the sample decreases monotonically with $I$ or that a real cooling has been achieved at the cold edge. This is because the Joule heating usually has a dominant contribution. The first and second harmonic *ac* temperature variations in response to a sine-wave *ac* current presented in Fig. 1 allow distinguishing between the Ettingshausen and Joule terms, but do not include the *dc* Joule heating component (Methods). To measure the actual map of excess temperature, we modulate the applied current between zero and $I$ by applying a unipolar square wave. In this case, the measured *ac* $\delta T(x,y)$ provides an image of the total current-induced excess temperature, which is the difference between the local temperature $T(x,y) = T_0 + \delta T(x,y)$ in the presence of the current and the base temperature $T_0$ in its absence. Positive $\delta T$ represents heating, while negative $\delta T$ is cooling to below $T_0$.

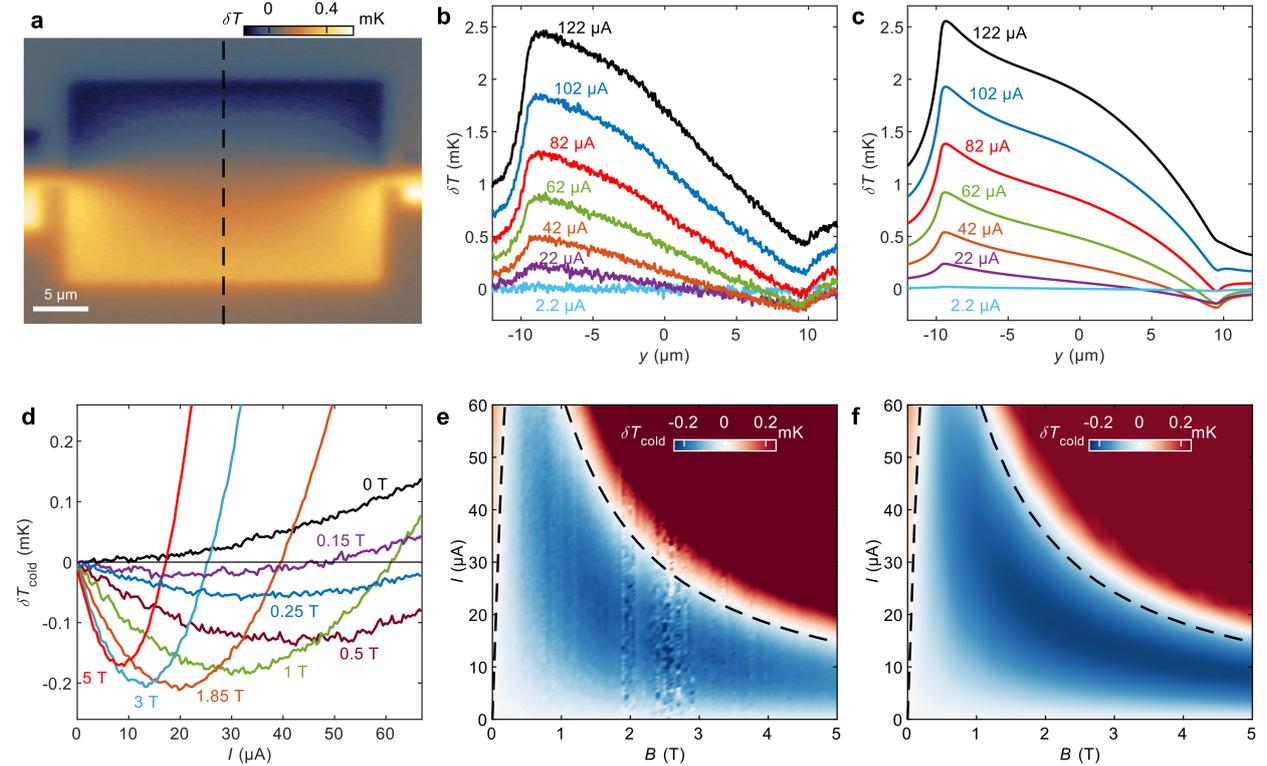

**Fig. 2. Absolute cooling.** (**a**) Map of the current-induced local excess temperature $\delta T(x,y)$ relative to the base temperature $T_0 = 4.3$ K, generated by unipolar square wave current excitation of $I = 25$ μA in WTe$_2$ device at $B = 0.64$ T. Regions with negative $\delta T$ are cooled by the current to below $T_0$. (**b**) Profiles of excess temperature $\delta T(0,y)$ along the black dashed line in (A) at various applied currents. (**c**) Simulated temperature profiles $\delta T(0,y)$. (**d**) Excess temperature at the cold edge of the chamber $\delta T_{cold}$ vs. $I$ at different magnetic fields showing absolute cooling $\delta T_{cold} < 0$. (**e**) The measured $\delta T_{cold}(I,B)$ diagram showing the parameter space of absolute cooling (blue). (**f**) Simulated $\delta T_{cold}(I,B)$ diagram. The color in the high-temperature regions (red) in (e,f) is saturated for clarity. Dashed line in (e,f) shows the fit of the current $I_0(B)$ from Eq. (7) below which absolute cooling occurs.

Figure 2a shows the resulting temperature map $\delta T(x,y)$ in the $W = 20$ μm chamber at $B = 0.64$ T and $I = 25$ μA. $\delta T(x,y)$ along the top edge of the sample is negative, namely the local temperature in the presence of current is below $T_0$, revealing current-induced absolute cooling. Even stronger cooling is observed in the top two corners of the chamber. To the best of our knowledge, this is the first observation of Ettingshausen cooling at liquid He temperatures.

Next we explore the current and field dependence of the cooling mechanism. Figure 2b shows the temperature profile $\delta T(0,y)$ across the sample center (dashed line in Fig. 2a) at $B = 0.64$ T and various applied currents $I$. The temperature at the hot edge ($y = -\frac{W}{2}$) increases monotonically with the current and the temperature difference $\Delta T_E$ between the hot and cold edges increases linearly with $I$ (Fig. 1j). Yet, the temperature at the cold edge ($y = \frac{W}{2}$) shows a nonmonotonic behavior due to competition between the linear Ettingshausen cooling and the quadratic Joule heating. As a result, absolute cooling ($\delta T < 0$) is achieved only at low currents. Note that at any current the temperature at the cold edge is lower than the temperature at the hot edge or in the sample center, namely the relative Ettingshausen cooling is always present, but at high currents it cannot overcome the Joule heating to attain absolute cooling.

The cooling mechanism is also nonmonotonic in magnetic field. Figure 2d shows $\delta T_{cold} = \delta T(0, \frac{W}{2})$ vs. $I$ at different $B$. At $B = 0$, the excess temperature is always positive, increasing monotonically with $I$. At finite fields, $\delta T_{cold}$ is negative at low currents, initially showing a negative slope with $I$ which increases monotonically with $B$. The current at which the maximal negative temperature $\delta T_{cold}^{max}(B)$ is attained decreases monotonically with $B$, but $\delta T_{cold}^{max}(B)$ has a nonmonotonic $B$ dependence with maximal negative $\delta T_{cold}^{max}$ attained at $B \cong 2$ T. Figure 2e shows the full map of $\delta T_{cold}(B, I)$ revealing a wide range of fields and currents at which negative excess temperatures can be achieved (blue).

**Theory of Ettingshausen effect in mesoscopic devices**

The mechanism underlying the large Ettingshausen effect in WTe$_2$ stems from the near compensation of electron and hole densities, $n_e \cong n_h = n$, and their mobilities, $\mu_e \cong \mu_h = \mu$. In bulk devices the induced transverse temperature difference, $\Delta T_E = P_E I B / W$, is proportional to $B/W$ [12,19,20]. This relation does not hold in mesoscopic devices because of the competition between different microscopic length scales and due to the inherently non-uniform current distribution. Consider a bulk sample with uniform applied longitudinal electric field $\boldsymbol{E} = E_x \hat{x}$. At $B = 0$ T, the resulting current density $\boldsymbol{J}$ comprises counterflowing electron $\boldsymbol{j}_e$ and hole $\boldsymbol{j}_h$ particle flux densities, $\boldsymbol{J} = e(\boldsymbol{j}_h - \boldsymbol{j}_e)$, where $e$ is the elementary charge. An out-of-plane magnetic field $\boldsymbol{B} = B\hat{z}$ gives rise to a transverse Lorentz force $\boldsymbol{F}_L$, which in the case of a single-charge-type conductor is counterbalanced by the electrostatic force of the induced Hall voltage, $\boldsymbol{F}_E$, resulting in absence of magnetoresistance (Fig. 1a). In a compensated semimetal, in contrast, the Hall voltage vanishes and $\boldsymbol{F}_L$, which scales as $\mu B$, drives both the electrons and holes in the same transverse direction (Fig. 1b). This transverse flow of quasiparticles, creates the quasiparticle current density $\boldsymbol{P} = e(\boldsymbol{j}_h + \boldsymbol{j}_e)$, which in turn causes a longitudinal Lorentz force counteracting the force of the applied electric field, resulting in bulk magnetoresistance that scales as $(\mu B)^2$, $\rho_{xx} \cong \rho_0(1 + \mu^2 B^2)$, where $\rho_0$ is the zero-field resistivity [29–32,51].

In a microscopic device, the edges play a crucial role. In particular, the boundary conditions preclude the transverse particle flow at the sample edges. This implies that the magnetoresistance, which is caused by the transverse flow, is suppressed near the boundaries. A recent theoretical study [52,53] shows that the resulting current distribution across the width $-\frac{W}{2} \leq y \leq \frac{W}{2}$ of a narrow long strip is given by

$$J_x(y) = \frac{E_x}{\rho_0(1 + \mu^2 B^2)} \left[1 + \mu^2 B^2 \frac{\cosh\left(\frac{2y}{l_R}\right)}{\cosh\left(\frac{W}{l_R}\right)}\right], \tag{1}$$

where $l_R = l_R^0/\sqrt{1+\mu^2 B^2}$ and $l_R^0$ is the electron-hole recombination length at zero field. For $l_R \ll W$, Eq. (1) shows that in the center of the strip the current density is suppressed by the magnetoresistance just like in the bulk limit, $J_x(|y| \ll W) = \frac{E_x}{\rho_0(1+\mu^2 B^2)}$. Remarkably, along the edges, the current experiences no magnetoresistance, $J_x\left(|y| = \frac{W}{2}\right) = \frac{E_x}{\rho_0}$, remaining at its zero-field value. As a result, the current distribution is peaked within narrow channels along the edges with characteristic width of $l_R$ (narrow bright slivers along the top and bottom edges in Fig. 3a).

In contrast to the conservation of charge current, $\nabla \cdot \boldsymbol{J} = 0$, the quasiparticle current $\boldsymbol{P}$ is not conserved because electron-hole pairs can be generated and recombined. Hence $\nabla \cdot \boldsymbol{P} = -\delta n_q/\tau_R$, where $\delta n_q = \delta n_e + \delta n_h$ is the quasiparticle excess density, $\delta n_e$ and $\delta n_h$ are the out-of-equilibrium electron and hole densities, and $\tau_R$ is the electron-hole recombination time. In the narrow strip geometry, the resulting quasiparticle current and density distributions are given by

$$P_y(y) = -\frac{\mu B E_x}{\rho_0(1+\mu^2 B^2)}\left[1 - \frac{\cosh\left(\frac{2y}{l_R}\right)}{\cosh\left(\frac{W}{l_R}\right)}\right]. \tag{2}$$

$$\delta n_q(y) = -\frac{2\mu B E_x}{\rho_0(1+\mu^2 B^2)}\frac{\tau_R}{l_R}\frac{\sinh\left(\frac{2y}{l_R}\right)}{\cosh\left(\frac{W}{l_R}\right)}. \tag{3}$$

In the central part of the sample, $\delta n_q$ is small because the quasiparticle thermal generation and recombination rates are approximately balanced like in thermal equilibrium, and $P_y$ is essentially constant as shown in Figs. 3b,c. The transverse $P_y$ driven by the Lorentz force gives rise to a sharp accumulation (depletion) of the quasiparticles within $l_R$ from the bottom (top) edge. As a result, the electron-hole pair recombination rate is enhanced (suppressed) by $\delta n_q/\tau_R$ at the bottom (top) edge, giving rise to heating by phonon emission (cooling by phonon absorption).

The thermoelectric effect can thus be derived by considering the two heat generating terms, $\dot{Q} = \dot{Q}_J + \dot{Q}_E$, where $\dot{Q}_J = \boldsymbol{J} \cdot \boldsymbol{E}$ is the Joule term, $\dot{Q}_E = \epsilon_p \delta n_q/\tau_R = -\epsilon_p \nabla \cdot \boldsymbol{P}$ is the Ettingshausen contribution, and $\epsilon_p$ is the average phonon energy released (absorbed) by recombination (generation) of an electron-hole pair:

$$\dot{Q}_J(y) = \frac{E_x^2}{\rho_0(1+\mu^2 B^2)}\left[1 + \mu^2 B^2 \frac{\cosh\left(\frac{2y}{l_R}\right)}{\cosh\left(\frac{W}{l_R}\right)}\right], \tag{4}$$

$$\dot{Q}_E(y) = -\frac{2\epsilon_p \mu B E_x}{e\rho_0 l_R^0 \sqrt{1+\mu^2 B^2}}\frac{\sinh\left(\frac{2y}{l_R}\right)}{\cosh\left(\frac{W}{l_R}\right)}. \tag{5}$$

As expected, $\dot{Q}_J$ is symmetric in $y$ and quadratic in $E_x$ and hence appears in the second harmonic for *ac* current, while $\dot{Q}_E$ is antisymmetric in $y$ and linear in $E_x$, consistent with the experimental results in Fig. 1.

To attain a better understanding of the thermoelectricity in mesoscopic systems, we perform 3D finite-element numerical simulations for our sample geometry (Methods). Figure 3a shows that a high current density $\boldsymbol{J}$ is present in the constrictions, while in the rectangular chamber the current density peaks along the sample edges where the magnetoresistance is suppressed. The corresponding Joule dissipation $\dot{Q}_J$ is highest in the constrictions (Fig. 3d) and is enhanced in the narrow slivers along the top and bottom edges in the chamber. The quasiparticle current density $\boldsymbol{P}(x,y)$ in Fig. 3b also shows a large enhancement in the constrictions. In the rectangular chamber $\boldsymbol{P}$ flows mostly transverse to $\boldsymbol{J}$ and is largest in the central area. Note that $|\boldsymbol{P}| \cong \mu B|\boldsymbol{J}|$ is substantially larger than $|\boldsymbol{J}|$ for the presented case of $\mu B = 1.6$ corresponding to our experimental values at $B = 0.64$ T. The excess quasiparticle density $\delta n_q(x,y)$ sharply peaks along the sample boundaries including the left and right edges (Fig. 3c). As a result, the Ettingshausen dissipation $\dot{Q}_E$ (Fig. 3e) shows pronounced cooling along the edges in the top half of the sample and heating along the bottom-half edges. Depending on the parameters, the total dissipation $\dot{Q} = \dot{Q}_J + \dot{Q}_E$ can be either positive everywhere, or negative (cooling) along the top boundaries as shown in Fig. 3f.

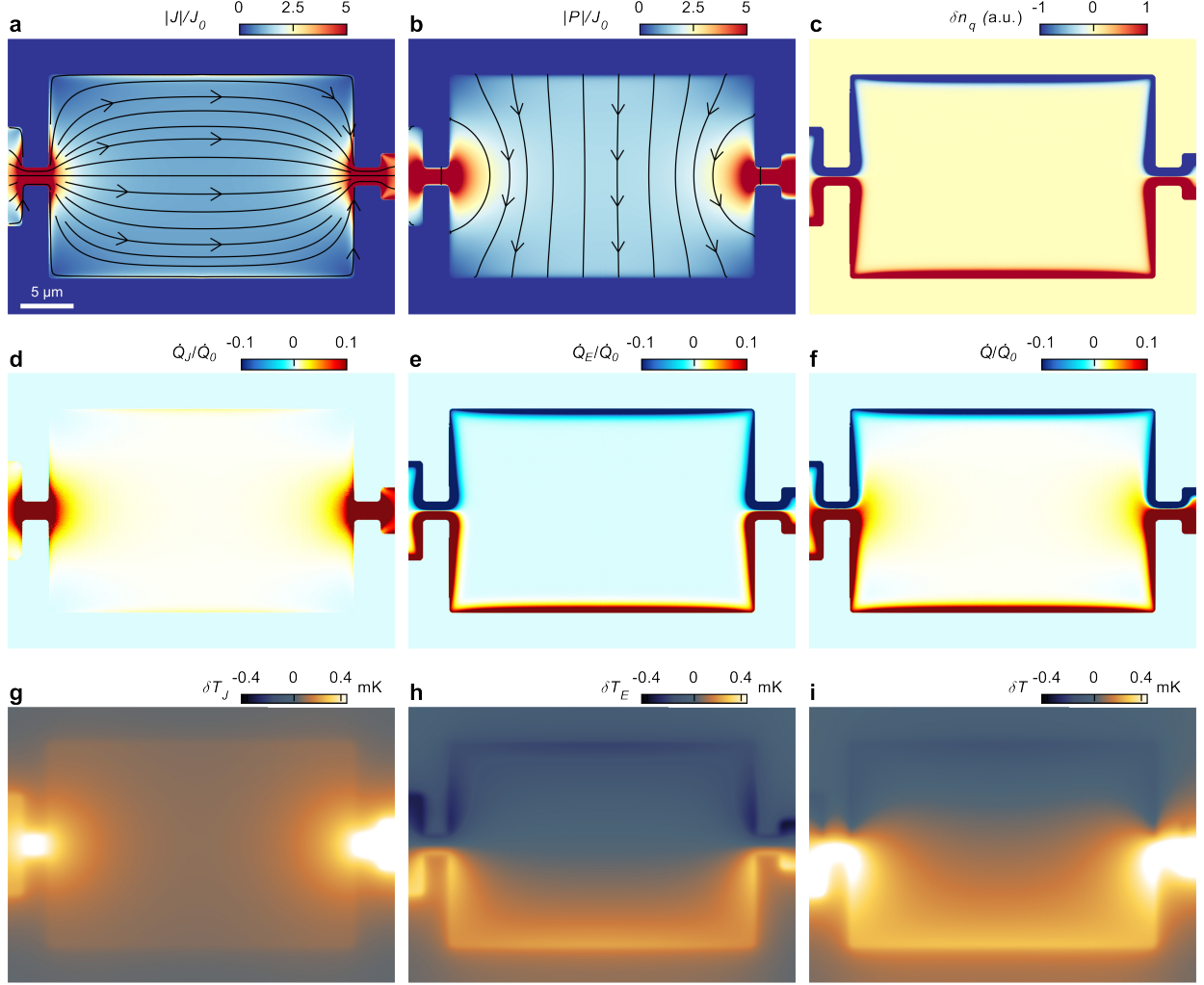

**Fig. 3. Numerical simulations of magneto-thermoelectric response**. (**a**) Normalized current density distribution $|J(x,y)|$ and streamlines at $B = 0.64$ T and $I = 25$ µA ($J_0 = I/W$). (**b**) Quasiparticle current density $|P(x,y)|$ with streamlines. (**c**) Out-of-equilibrium quasiparticle density $\delta n_q(x,y)$. The colors are saturated for clarity. (**d**) Power density generated by Joule heating $\dot{Q}_J(x,y)$. (**e**) Power density $\dot{Q}_E(x,y)$ produced by generation and recombination of quasiparticles due to Ettingshausen effect. (**f**) Total power density $\dot{Q} = \dot{Q}_J + \dot{Q}_E$. (**g**) Excess temperature distribution $\delta T_J$ caused by Joule heating. (**h**) Excess temperature distribution $\delta T_E$ caused by Ettingshausen effect. (**i**) Total excess temperature $\delta T(x,y)$. See Methods for simulation parameters.

By incorporating heat diffusion equations in WTe$_2$ and in the Si substrate into the numerical calculations (Methods), we can derive the corresponding $\delta T_J$, $\delta T_E$, and the total $\delta T$ distributions. Figures 3g-i show that the calculated local temperatures follow the local $\dot{Q}$ maps in Figs. 3d-f, albeit broadened by the thermal heat conductivities in the flake and the substrate. In particular, the enhancement of $\dot{Q}_J$ in the narrow slivers along the edges in Fig. 3d is completely broadened in the $\delta T_J$ map in Fig. 3g. We attain very good agreement between the numerical simulations and the experimental data for the microscopic parameters $l_R^0 = 0.5$ µm, $\mu = 25{,}000$ cm$^2$V$^{-1}$s$^{-1}$, and $\epsilon_p = 0.06$ meV. The temperature distributions calculated in Fig. 3g-i match qualitatively very well the measured $\delta T_J$ and $\delta T_E$ maps in Figs. 1e,h and with the total $\delta T$ distribution in Fig. 2a. Figure 2c show the calculated $\delta T(y)$ profiles across the width of the sample at various currents, well reproducing the experimental behavior. Moreover, the experimental phase diagram of the absolute cooling $\delta T_{cold}(I, B)$ in Fig. 2e is also well described by the model (Fig. 2f), showing the accessible range of absolute cooling and the optimal parameters for achieving the lowest temperatures in our device. Finally, the model also recovers the nonmonotonic width dependence and the nonlinearity of $\Delta T_E(B)$ in Fig. 1i (solid lines).

**Analytical model**

Further insight into the role of the different parameters in a mesoscopic device can be attained by considering a simplified model for the total heat fluxes. The lateral heat flux from the hot to the cold side of the device is governed by the in-plane total heat conductivity $\kappa$ of WTe$_2$ with thickness $d$. Further, heat flows from the hot side of the device to the substrate, and from the substrate to the cold part of the device, which depends on the out-of-plane heat conductivity $\kappa'$ of the substrate with effective thickness $d_s$. In this geometry, we find (Methods) that the condition on the applied current for attaining a negative excess temperature at the cold edge is $I < I_0$, where

$$I_0 \cong \frac{\epsilon_p}{e\rho_0} \frac{4\mu B}{(1 + \mu^2 B^2)(1 + W_0^2/W^2)} \tag{6}$$

and $W_0 \cong \sqrt{8 d d_s \kappa / \kappa'}$. The dashed curve in Fig. 2e shows the fit of $I_0(B)$ to the experimental data providing a very good description of the cooling regime. Note that from Eq. (6) the maximum of $I_0(B)$ occurs at $\mu B = 1$, which allows a direct evaluation of $\mu \cong 23{,}000$ cm$^2$V$^{-1}$s$^{-1}$ from the experimental data, consistent with the more accurate 3D simulations resulting in $\mu = 25{,}000$ cm$^2$V$^{-1}$s$^{-1}$.

By integrating $\dot{Q}_E$ over the cold side of the sample and taking into account the above heat conductivities, we derive (Methods) the transverse temperature difference

$$\Delta T_E \cong \frac{\epsilon_p I}{e d \kappa} \frac{2\mu B}{\left(1 + \frac{\mu B l_R^0}{W}\right)\left(1 + \frac{W^2}{W_0^2}\right)}. \tag{7}$$

This result emphasizes a number of key aspects of the mesoscopic nature of the Ettingshausen effect, which cannot be observed in bulk materials. The first is the sublinear dependence of $\Delta T_E$ on $B$ due to the $\frac{\mu B l_R^0}{W}$ term, which allows us to evaluate the electron-hole recombination length of $l_R^0 \cong 0.3$ μm, similar to the more accurate value derived numerically (Fig. 1l). In contrast, in the bulk limit of $W \gg \mu B l_R^0$, the common linear $B$ dependence of $\Delta T_E$ is recovered, losing the ability to determine $l_R^0$ from such measurements.

Secondly, Eq. (7) predicts a nonmonotonic dependence of $\Delta T_E$ on $W$, which describes well the behavior in Fig. 1l (see Extended Data Fig. 1 for more details). The largest $\Delta T_E$ should be attainable for $W \cong \sqrt[3]{W_0^2 l_R^0} \mu B$ for $\mu B < 1$. From our observation of the largest transverse temperature difference in the $W = 10$ μm sample, we derive $W_0 \cong 24$ μm. This new mesoscopic length scale $W_0$ arises due to the competition between the transverse heat conductivity across the sample width and the heat conductivity to the substrate, and is absent in the bulk case. This length scale can serve as an important parameter for further optimization of the Ettingshausen cooling in microscopic devices.

Third, we derive that the largest cooling is attained at $\hat{I} = I_0/2$ and $\hat{B} = \frac{1}{\mu}\sqrt[3]{\frac{2W}{l_R^0}}$ (Methods) where the negative excess temperature reaches

$$\delta T_{cold}^{max} \cong -\frac{\epsilon_p^2}{e^2 \rho_0 d \kappa} \frac{W_0^2 W^2}{(W^2 + W_0^2)^2}. \tag{8}$$

For $W = 20$ μm this results in $\hat{B} \cong 2.2$ T and $\delta T_{cold}^{max} \cong -0.2$ mK in good agreement with Fig. 2d.

Equation 8 shows that lower cooling temperatures can be attained by reducing the sample sheet resistivity $\rho_0$. In our WTe$_2$ flake $\rho_0 \cong 0.54$ Ω, corresponding to effective bulk resistivity of $\rho_0 d \cong 1.5 \times 10^{-5}$ Ω·cm ($d = 269$ nm), which is almost two orders of magnitude higher than the low-temperature bulk resistivity, $\rho \cong 2.3 \times 10^{-7}$ Ω·cm, in high quality crystals [31]. Since the increase in $\rho_0$ is attributed to surface oxidation [54], one can expect attaining significantly stronger cooling in devices fabricated in inert environment and encapsulated in hBN.

Finally, from the combination of the derived parameters (Methods) we extract the Ettingshausen conversion coefficient $\epsilon_p \cong 0.06$ meV, which is about one sixth of the thermal energy of the quasiparticles ($k_B T_0$ = 0.37 meV), and thus in good agreement with the description of the Ettingshausen effect in terms of particle-hole generation and recombination.

The developed nanoscale imaging of magneto-electro-thermal cooling at cryogenic temperatures in microscopic devices opens the door for study and engineering of thermal landscapes and energy harvesting in novel electronic materials and devices. With this technique, we have demonstrated absolute cooling over a wide range of magnetic fields and applied currents and found an ultrahigh Ettingshausen effect at liquid He temperature in exfoliated flakes of Weyl semimetal WTe$_2$. The derived microscopic model allows one to further enhance the cooling efficiency through the optimization of the electrical and thermal conductivity characteristics of the materials and of the device geometries. This study suggests the integration of transition metal dichalcogenide atomic layer semimetals in multilayer van der Waals heterostructures to provide for in-device electro-thermal cooling, temperature gradient engineering, and study of thermoelectric effects in low-dimensional strongly correlated states of matter.

**Acknowledgements**

**Funding:**

European Union (ERC, MoireMultiProbe - 101089714). "Views and opinions expressed are however those of the author(s) only and do not necessarily reflect those of the European Union or the European Research Council. Neither the European Union nor the granting authority can be held responsible for them." (EZ).
Israel Science Foundation ISF grant No 687/22 (EZ, MH).
GIF German-Israeli Foundation for Scientific Research and Development (GIF) Grant no. I-1505-303.10/2019 (EZ).
Goldfield Family Charitable Trust (EZ).
Andre Deloro Prize for Scientific Research (EZ).
Leona M. and Harry B. Helmsley Charitable Trust grant 2112-04911 (EZ, MH).

**Author contributions:**

Cryogenic thermal imaging measurements: TV, AA-S
Growth and characterization of $WTe_2$ crystals: AKP, MH
Sample design and fabrication: EA
SOT fabrication and tuning fork feedback: NB, YM
Design and building of the SOT readout system: MEH
Finite-element COMSOL numerical simulations: AA-S
Development of analytical derivation of Ettingshausen effect: TH AA-S
Data analysis: TV, AA-S, EA
Writing of original manuscript: TV, EZ, AA-S, TH
Editing and review of manuscript: all authors

**Competing interests:**

The authors declare no competing interests.

**Data and materials availability:**

The data that support the findings of this study are available from the corresponding author on reasonable request. The finite-element COMSOL numerical simulation codes are available from the corresponding author on reasonable request.

## Methods

**WTe₂ device fabrication**

High quality WTe$_2$ single crystals were grown using a flux growth technique [51] as described in [31]. Bulk samples produced by this method showed a residual resistance ratio of up to $RRR = 3250$, magnetoresistance ratio of up to 62,000 at 9 T and 2 K, and a mean free path of $l_{mr} \cong 20$ μm [31]. The crystals were then used to mechanically exfoliate WTe$_2$ flakes onto oxidized silicon wafers (290 nm of SiO$_2$). Suitable flakes, identified by optical and atomic force microscopy, were then processed by electron beam lithography (EBL) and inductively coupled plasma etching (ICP) into the various geometries. ICP was done with a flow of 20 sccm of SF$_6$ and 10 sccm of O$_2$, at radio-frequency power of 25 W, which provides an etching rate of WTe$_2$ of about 4.7 nm per minute. Electrical contacts were fabricated by an additional EBL step, Ar ion milling to remove the native oxide layer of WTe$_2$, followed by E-beam deposition of 3 nm of Ti and Au of a thickness that exceeded the WTe$_2$ flake thickness. Transport measurements of the flakes show carrier mobilities of about 10,000 to 30,000 cm$^2$V$^{-1}$s$^{-1}$ at 4 K, corresponding to mean free path of $l_{mr} \cong 0.7$ to 2 μm, more than an order of magnitude lower than the bulk values due to surface oxidation and scattering as reported previously [31,54].

**SOT fabrication and thermal imaging**

Nanoscale SOTs have been used as high precision local thermometers to study dissipation in mesoscopic systems [50,55–57]. This is achieved through the strong dependence of the SOT critical current $I_c(T)$ on the SOT temperature. In practice, the SOT is connected in parallel with a small shunt resistor and biased above the critical current, $I_{bias} > I_c$. As a result, the current through the SOT, $I_{SOT}(T)$, has temperature dependence that approximately follows $I_c(T)$, displaying best sensitivity in the temperature range $T_c/2 < T < T_c$. The thermal coupling between the SOT and the sample is achieved by a low pressure of He exchange gas. A cryogenic SQUID series array amplifier (SSAA) [58] is used for the $I_{SOT}$ readout.

SOTs with diameter around 110 nm were fabricated as described in [48,59]. In this work, MoRe ($T_c \cong 7.2$ K) was used as the superconducting material for the SOT to provide thermal sensitivity throughout the magnetic field range up to 5 T [49] with a thermal sensitivity at zero field of 2.6 μK/Hz$^{1/2}$. Calibration of temperature sensitivity of the SOT at zero field was carried out using a heater installed in the microscope, and the relative magnetic field dependence of the sensitivity was derived by imaging the temperature of an Au thin-film heater patterned next the WTe$_2$ flake vs. $B$. The MoRe SOT was intentionally chosen to have a weak magnetic field sensitivity which decreased rapidly with magnetic field [49]. All the presented data were acquired under conditions that the magnetic signal due to the Oersted field of the current flowing in the sample is negligible as compared to the thermal signal, except for Fig. 1g, where a small signal at zero field can be resolved in the constrictions.

The SOT height control was achieved by attaching the tip to a quartz tuning fork [50], which allowed scanning at a height of 80 nm above the sample surface. Thermal imaging was done at base temperature of 4.3 K and in an environment of 40 mbar of He exchange gas to provide the thermal link between the tip and the sample [50]. All images in the main text were acquired with image resolution of 344×269 pixels, pixel size of 90 nm, and acquisition time of 40 ms per pixel.

For the scans in Fig. 1, an *ac* bipolar sine wave with rms current of $I = 50$ μA and excitation frequency of $f = 85.37$ Hz was applied, and the resulting thermal response of the SOT was measured at the 1$^{st}$ and 2$^{nd}$ harmonics of $f$ using a lock-in amplifier. Note that a current $I(t) = I_0 \cos(\omega t)$ creates Joule heating $\dot{Q}_J(t) = RI^2(t) = \frac{1}{2}RI_0^2 + \frac{1}{2}RI_0^2 \cos(2\omega t)$, which has a *dc* component and a superimposed *ac* component at twice the excitation frequency. The latter *ac* component is measured by the 2$^{nd}$ harmonic signal of the SOT.

In order to capture the total current-induced local temperature change in the sample, $\delta T$, relative to the base temperature, $T_0$, we apply a unipolar square wave excitation and measure the difference between the current-on and the current-off states as presented in Fig. 2a. Negative $\delta T$ means absolute current-induced cooling, namely in the current-on state the local temperature is lower than in the current-off state.

**Analytical derivation of the Ettingshausen effect**

Here, we provide the details of the simplified heat flux model, which is in good agreement with the full numerical simulations and provides a deeper insight into the role of the different parameters. We assume an approximately linear temperature gradient across the width $-\frac{W}{2} \leq y \leq \frac{W}{2}$ of the sample, $T(y) = T_0 - |\delta T_{cold}| + (0.5 + y/W)\Delta T_E$, and calculate the lateral and out-of-plane heat fluxes. The lateral heat flux through the flake from the hot to the cold side of the device is $q_0 = d\kappa \Delta T_E/W$. We compare the heat fluxes into and out of the substrate by separately considering the two regions of the sample, $-\frac{W}{2} \leq y \leq 0$ and $0 < y \leq \frac{W}{2}$. By integrating over each region, we attain the heat flux from the hotter part of the sample into the substrate, $q_h = \frac{\kappa'}{d_s} \int_0^{W/2} (T(y) - T_0)dy = W\kappa'(3\Delta T_E - 4|\delta T_{cold}|)/8d_s$, and similarly, the heat flux from the substrate into the colder side by $q_c = \frac{\kappa'}{d_s} \int_{-W/2}^{0} (T_0 - T(y))dy = W\kappa'(4|\delta T_{cold}| - \Delta T_E)/8d_s$. This leads to $q_h + q_c = W\kappa'\Delta T_E/4d_s$ and $q_h - q_c = W\kappa'(\Delta T_E - 2|\delta T_{cold}|)/2d_s$. The total heat flow generated by the Ettingshausen heating is thus described by

$$2\int_{-W/2}^{0} \dot{Q}_E dy \cong 2q_0 + q_h + q_c = \left(\frac{2d\kappa}{W} + \frac{W\kappa'}{4d_s}\right)\Delta T_E, \tag{9}$$

while the heat flow due to Joule heating in the same area is given by

$$2\int_{-W/2}^{0} \dot{Q}_J dy = q_h - q_c = \frac{W\kappa'}{2d_s}(\Delta T_E - 2|\delta T_{cold}|). \tag{10}$$

Introducing the length scale $W_0 = \sqrt{8dd_s \frac{\kappa}{\kappa'}}$ and solving for $|\delta T_{cold}|$ and the temperature difference $\Delta T_E$, thus results in

$$\Delta T_E = \frac{\epsilon_p I}{ed\kappa} \frac{2\mu B}{(1 + \mu B l_R^0/W)(1 + W^2/W_0^2)}, \tag{11}$$

$$|\delta T_{cold}| = \frac{\frac{\rho_0 I^2}{4}(1 + \mu^2 B^2) - I \frac{\epsilon_p}{e} \frac{\mu B}{1 + W_0^2/W^2}}{d\kappa \frac{W^2}{W_0^2}\left(1 + \frac{\mu^2 B^2}{\sqrt{1 + \mu^2 B^2}} \frac{l_R^0}{W}\right)}. \tag{12}$$

The current at which the Joule heating at the cold edge balances the Ettingshausen cooling, resulting in $|\delta T_{cold}| = 0$, is thus given by

$$I_0 \cong \frac{\epsilon_p}{e\rho_0} \frac{4\mu B}{(1 + \mu^2 B^2)\left(1 + \frac{W_0^2}{W^2}\right)}. \tag{13}$$

The maximal cooling is attained at $\hat{I} = I_0/2$ and $\hat{B} = \frac{1}{\mu}\sqrt[3]{\frac{2W}{l_R^0}}$, with the lowest temperature of

$$\delta T_{cold}^{max}(\hat{B}) \cong -\frac{\epsilon_p^2}{e^2 \rho_0 d\kappa} \frac{W_0^2 W^2}{(W^2 + W_0^2)^2}. \tag{14}$$

The fit to the experimental data is performed in three steps. First, the value of $\mu$ and the combined coefficient $C_1 = \frac{\epsilon_p}{e\rho_0} \frac{1}{(1+W_0^2/W^2)}$ can be determined from fitting the line $|\delta T_{cold}| = 0$ in the phase diagram in Fig. 2e. Subsequently, the nonmonotonic dependence of $\Delta T_E$ as a function of $W$ for a given current $I$ in Eq. 11 can be used to yield an estimate for $W_0$, $l_R^0$ and the global coefficient $C_2 = \frac{\epsilon_p}{ed\kappa}$, as shown in Extended Data Fig. 1a,b. These values can then be used to evaluate $\delta T_{cold}^{max} = -C_1 C_2 \frac{W_0^2}{(W^2+W_0^2)}$ (Extended Data Fig. 1c). Finally, from $C_1$ and using the zero-field sheet resistance $\rho_0 \cong 0.54$ Ω derived from the full numerical fits, we can evaluate $\epsilon_p \cong 0.06$ meV.

Extended Data Figs. 1 and 2e,f show that the analytical derivation captures well the essence of the Ettingshausen effect and of the absolute cooling in mesoscopic devices, although it considers a simplified model compared with the numerical simulations presented in Figures 1l and 2f. The fitted values of $\mu$ and $l_R^0$ are slightly different between the analytical fits and the numerical simulations.

**Numerical simulations**

Finite-element 3D numerical simulations of the temperature maps due to Joule heating and Ettingshausen effect in a nearly compensated semimetal were used to fit the experimental data. The calculations were conducted using COMSOL Multiphysics 5.4 in two consecutive steps. In the first step, the transport equations of a two-component conductor [53] were solved:

$$\frac{l_R^2}{\tau_R} \nabla \delta n_\alpha - n_\alpha \mu_\alpha \boldsymbol{E} - \boldsymbol{j}_\alpha \times (\mu_\alpha \boldsymbol{B}) = -\boldsymbol{j}_\alpha \tag{15a}$$

$$\text{div } \boldsymbol{j}_{e(h)} = -\frac{\delta n_q}{2\tau_R} \tag{15b}$$

where the index $\alpha = e, h$ describes electrons or holes, $\boldsymbol{j}_\alpha$ are the particle flux densities, $\boldsymbol{E} = -\nabla\phi$ is the electric field, $\phi$ is the electric potential, $n_e$ and $n_h$ is the equilibrium density of electrons and holes, $\delta n_q = \delta n_e + \delta n_h$ is the total out-of-equilibrium quasiparticle density, $\mu_{e(h)}$ is the electron (hole) mobility, $\tau_R$ is the characteristic recombination time and $l_R$ is the recombination length. The recombination length and time are related by $l_R = \sqrt{D\tau_R}$, where $D$ is the diffusion coefficient. To represent these equations, we used the Coefficients Form PDE module, which solves the general equation

$$e_a \frac{\partial^2 \boldsymbol{u}}{\partial t^2} + d_a \frac{\partial \boldsymbol{u}}{\partial t} + \nabla \cdot (-c\nabla\boldsymbol{u} - \alpha\boldsymbol{u} + \gamma) + \beta \cdot \nabla\boldsymbol{u} + a\boldsymbol{u} = f, \tag{16}$$

where the field $\boldsymbol{u}$ is

$$\boldsymbol{u} = \begin{pmatrix} \tilde{\phi} & \widetilde{\delta n_q} & ej_{e,x} & ej_{e,y} & ej_{h,x} & ej_{h,y} \end{pmatrix}. \tag{17}$$

Here $\tilde{\phi} = \left(\frac{e^2}{h}\right)\frac{1}{ld}\phi$, $\widetilde{\delta n_q} = \frac{l_R}{\tau_R} e \delta n_q$, and $l = 1$ μm is a length scale for expressing $\boldsymbol{u}$ in units of current density. Equation (16) was solved in a 3D geometry that follows the dimensions of the experimental sample depicted in Fig. 1c, using fitted parameters $\mu_e = \mu_h = 25{,}000$ cm²V⁻¹s⁻¹, $n_e = n_h = 2.1 \cdot 10^{18}$ cm⁻³, and $l_R^0 = 0.5$ μm.

In the second step of the simulation the heat transport was solved. For this, we employed a heat transfer in solids module to simulate heat diffusion in the system. The 3D geometry consisted of the WTe₂ sample of thickness $d = 269$ nm, heat conductivity of $\kappa_{WTe_2} = 6.4$ W/K·m, a He exchange gas layer surrounding the sample and extends 1 μm above ($\kappa_{He} = 0.005$ W/K·m), a SiO₂ layer of thickness 300 nm ($\kappa_{SiO_2} = 0.2$ W/K·m), and a Si layer of thickness 5 μm ($\kappa_{Si} = 0.32$ W/K·m), which was taken for reasons of practical convenience. From the solution of Eq. (16) in the first step, we calculated $\dot{Q}_E = \epsilon_p \delta n_q/\tau_R$, $\dot{Q}_J = -\boldsymbol{J} \cdot \nabla\phi$, and $\dot{Q} = \dot{Q}_J + \dot{Q}_E$ with $\epsilon_p = 0.16 k_B T_0 = 0.06$ meV, and used these as the heat sources for the heat transfer module to solve for $\delta T_E$, $\delta T_J$ and $\delta T$ respectively. The bottom side of the Si layer was set as the thermal anchor to $T_0 = 4.3$ K.

Note that while $l_R^0$ and $\mu$ affect the fitting to the experimental data substantially, the rest of the parameters provide a fit over a range of values. The presented simulations should thus be considered as a qualitative demonstration of the validity of the model, rather than accurate determination of the parameter's values.

**Ettingshausen effect in additional samples**

**Strip geometry.** To exclude excess heating arising from the narrow constrictions in the rectangular chamber geometry, a sample with a strip geometry (Extended Data Fig. 2a) of width $W = 20$ µm and thickness $d = 160$ nm was investigated. By applying a sinusoidal *ac* current with $I = 50$ µA at $B = 5$ T, we acquired the maps of $\delta T_J$ and $\delta T_E$ shown in Extended Data Fig. 2d,e. The absence of the constrictions results in more uniform temperature distributions in the device as compared to Figs. 1f,i. The slight canting of the thermal distributions in Extended Data Fig. 2d,e is likely due to the small Hall-voltage resulting from incomplete compensation between the electron and hole densities and mobilities and shows an opposite tilt angle upon reversing the magnetic field direction (not shown). Quantitative analysis of the field dependence of $\delta T_J$ at the center of the device (Extended Data Fig. 2b) shows quadratic behavior that approximately follows the $B$ dependence of the sample resistivity, while the temperature difference $\Delta T_E = \delta T_E \left(0, -\frac{W}{2}\right) - \delta T_E \left(0, \frac{W}{2}\right)$ between the hot and cold edges across the center of the strip shows approximately linear $B$ dependence consistent with Figs. 1k,l.

Applying unipolar square wave excitation with $I = 8$ µA at $B = 5$ T reveals absolute cooling at the top edge in Extended Data Fig. 2f. The phase space for absolute cooling is found by mapping $\delta T_{cold}$ vs. $B$ and $I$ as shown in Extended Data Fig. 2c. The overall behavior is similar to the $W = 20$ µm device in Fig. 2e with comparable $\delta T_{cold}$ values. The main two quantitative differences, however, are the position of the peak of $I_0(B)$, which occurs at $B \cong 1.1$ T in Extended Data Fig. 2c and the maximal negative $\delta T_{cold}^{max}$ attained at $B \gtrsim 5$ T. By fitting the $I_0(B)$ (dashed line in Extended Data Fig. 2c) we find $\mu = 9,000$ cm$^2$V$^{-1}$s$^{-1}$. This lower value is consistent with the fact that this device is thinner [31,54].

**Rectangular geometry.** We present here thermal imaging of additional two chambers of width $W = 10$ and 5 µm connected in series with the $W = 20$ µm chamber described in the main text (Fig. 1c). Extended Data Figs. 3a-c show $\delta T_J$ maps at $B = 0$, 1, and 5 T in the $W = 10$ µm chamber at $I = 50$ µA with corresponding $\delta T_E$ maps in Extended Data Figs. 3d-f. A similar set of data is presented in Extended Data Figs. 3g-l for the $W = 5$ µm chamber. The qualitative behavior is similar to the $W = 20$ µm chamber, however, $\Delta T_E$ shows a nonmonotonic dependence on $W$ as described in Fig. 1l.

**Additional geometries.** Additional samples with more complex geometries are presented in Extended Data Fig. 4. Since in this set of experiments an accurate calibration of the SOT thermal response was not available, the data are presented in relative units.

Extended Data Fig. 4a,d show a sample with thickness $d = 110$ nm that allows application of current through channels of different widths and along different directions. For the current path shown in Extended Data Fig. 4a, largest $\delta T_J$ is observed in Extended Data Fig. 4b in the two narrowest channels on the left and right sides. In contrast, the largest transverse $\Delta T_E$ occurs in the widest top-left channel in Extended Data Fig. 4c. Additionally, strongly nonlocal Ettingshausen heating and cooling can be observed in the vicinity of the right-angle corners along the current path as discussed below.

Application of the current along the central vertical channel in Extended Data Fig. 4d reveals that $\Delta T_E$ in the vertical strip sections in Extended Data Fig. 4f is smaller than the Ettingshausen heating and cooling spots appearing at the junctions with the horizontal channels. The temperature maps for the two different current paths in the sample illustrate the mesoscopic nature of the Ettingshausen effect, as discussed in the main text, along with its nonlocal character. The Joule heating is local in its nature because $\dot{Q}_J$ is mainly determined by the local $J$. In contrast, $\dot{Q}_E$ is not determined by the local $P$, but rather by $\nabla \cdot \boldsymbol{P}$.

These features are further corroborated by thermal imaging of the sample shown in Extended Data Fig. 4g with thickness $d = 48$ nm, which was used in [31]. It consists of a very narrow strip of $W = 0.55$ μm connected to two circular disks of 1.8 μm diameter. Extended Data Figs. 4h,i show that along the narrow horizontal strip the $\delta T_J$ is largest, while $\Delta T_E$ is vanishingly small. In narrow channels, with width comparable to $l_R$, the regions of electron-hole generation and recombination partially overlap and the induced Ettingshausen temperature gradient is greatly suppressed by the large transverse heat conductivity across the narrow channel. In sharp contrast, Extended Data Fig. 4i shows a very pronounced Ettingshausen heating and cooling in the circular chambers, where almost no current flows, emphasizing its nonlocal character.

Qualitatively, the dependence of $\delta T_E$ on the channel width can also be seen for the temperature maps of the sample depicted in Extended Data Fig. 4j,m with two connected channels of different width, with thickness of $d = 105$ nm. As seen in Extended Data Fig. 4l, the temperature difference between the hot and cold edges, $\Delta T_E$, is larger in the wider channel and there is a pronounced increase in Ettingshausen heating and cooling at the corners, similarly to the results in the main text Figs. 1h,i. As expected for the Ettingshausen effect, $\delta T_E$ reverses sign upon reversing the magnetic field direction in Extended Data Fig. 4o, while the Joule heating shown in Extended Data Fig. 4k,n remains unaffected.

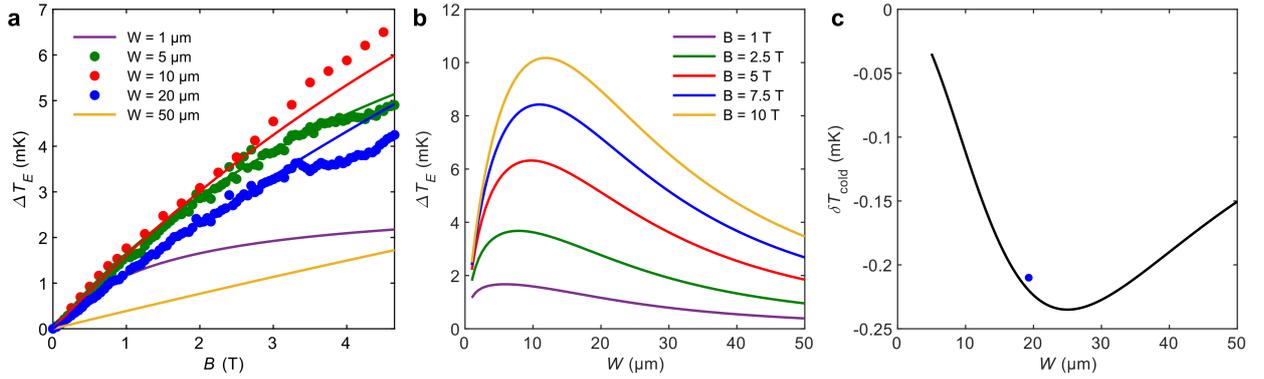

**Extended Data Fig. 1. Analytical calculations of Ettingshausen effect and absolute cooling in mesoscopic devices.** (**a**) Calculated $\Delta T_E$ vs. $B$ using Eq. 11 (solid lines), fitted to the experimental data from Fig. 1l for different widths $W$. The calculated $\Delta T_E$ curves show the nonmonotonic dependence on $W$ and the sublinear $B$ dependence for small $W$, which becomes linear for large $W$. The analytical curves, based on a simplified model, are slightly different from the numerical ones presented in Fig. 1l. (**b**) Calculated $\Delta T_E$ vs. $W$ for different magnetic field values $B$ showing nonmonotonic $W$ dependence. The width $W$ for which $\Delta T_E$ is largest grows monotonically with $B$. (**c**) Calculated $\delta T_{cold}^{max}$ (black line) showing a nonmonotonic dependence on $W$, peaked at $W_0$. The blue dot is the measured extremal $\delta T_{cold}$ from Fig. 2e.

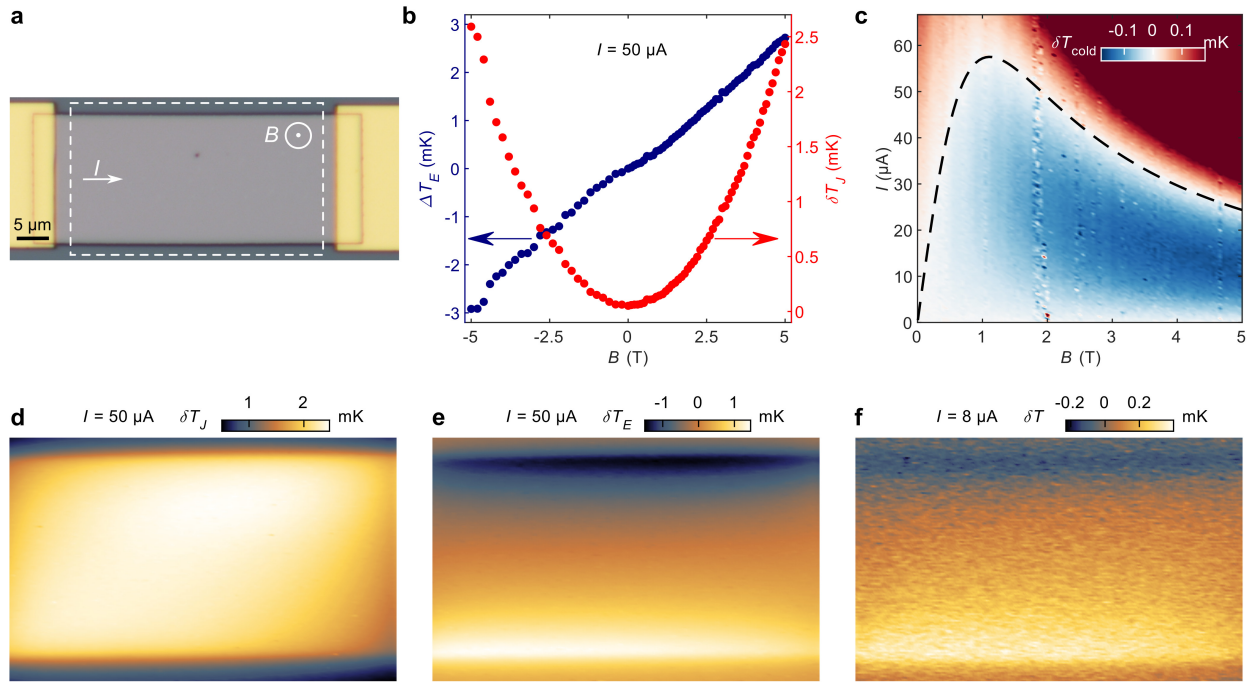

**Extended Data Fig. 2. Thermal imaging and absolute cooling in a sample with strip geometry.** (**a**) Optical image of the sample with length $L = 42$ μm and width $W = 20$ μm with Au contacts on the left and right sides. (**b**) Temperature difference $\Delta T_E$ between the bottom and the top edge (blue curve) and $\delta T_J$ in the center of the strip (red) vs. $B$ at *ac* current of 50 μA. (**c**) The temperature of the cold edge, $\delta T_{cold}$, as a function of $B$ and the applied unipolar current $I$. Dashed line shows a fit to the analytically calculated $I_0(B)$. (**d**) Thermal map of $\delta T_J$ at $B = 5$ T and $I = 50$ μA. (**e**) Corresponding map of $\delta T_E$. (**f**) Map of the total excess temperature $\delta T$ using unipolar square wave current $I = 8$ μA at $B = 5$ T, revealing absolute cooling at the top edge.

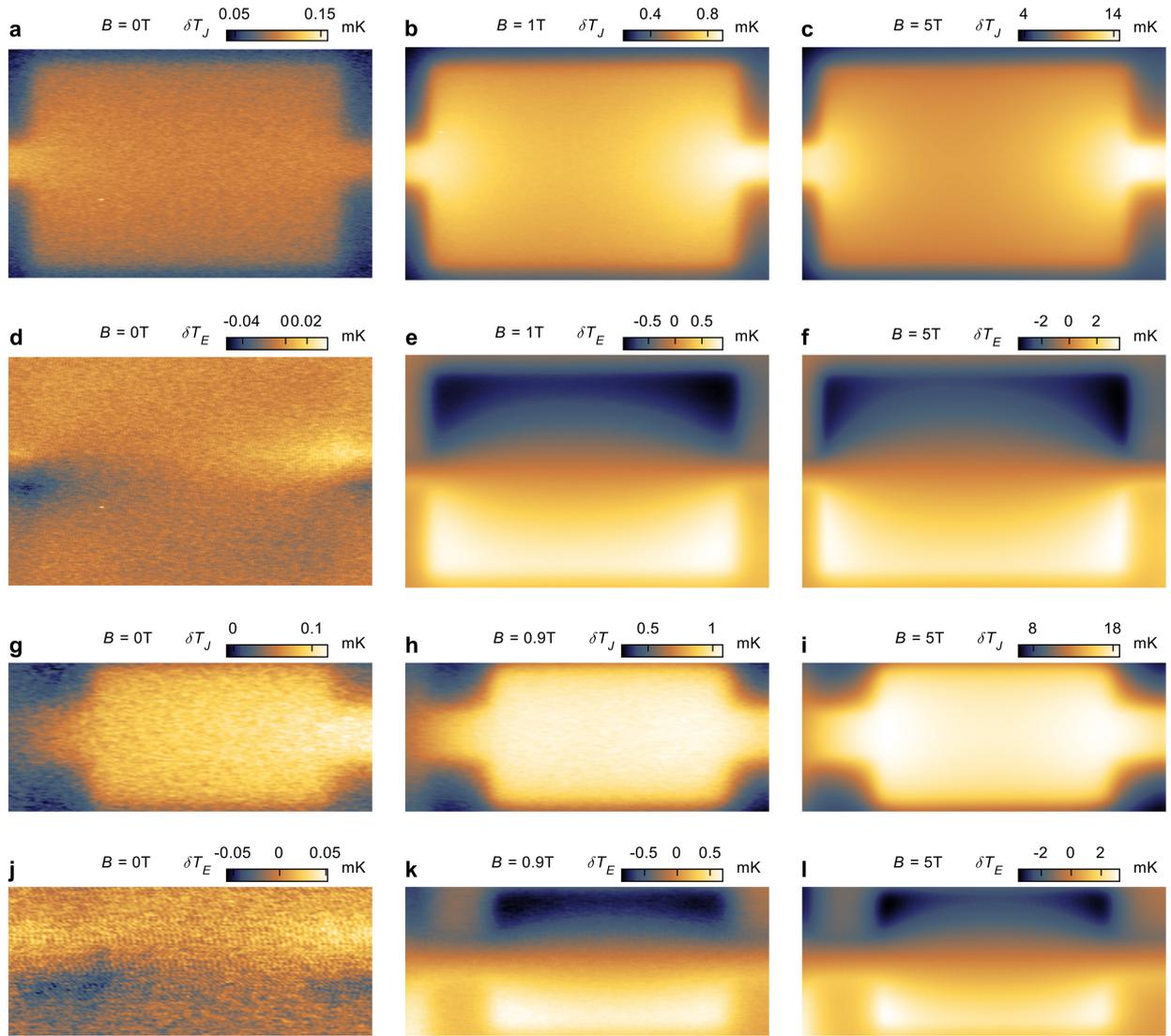

**Extended Data Fig. 3. Thermal imaging of additional chambers in rectangular geometry.** (**a-c**) Maps of $\delta T_J$ at different indicated magnetic field values for the rectangular chamber with width $W = 10$ µm at $I = 50$ µA. (**d-e**) Corresponding maps of $\delta T_E$. (**g-i**) Maps of $\delta T_J$ for the chamber with width $W = 5$ µm. (**j-l**) Corresponding maps of $\delta T_E$.

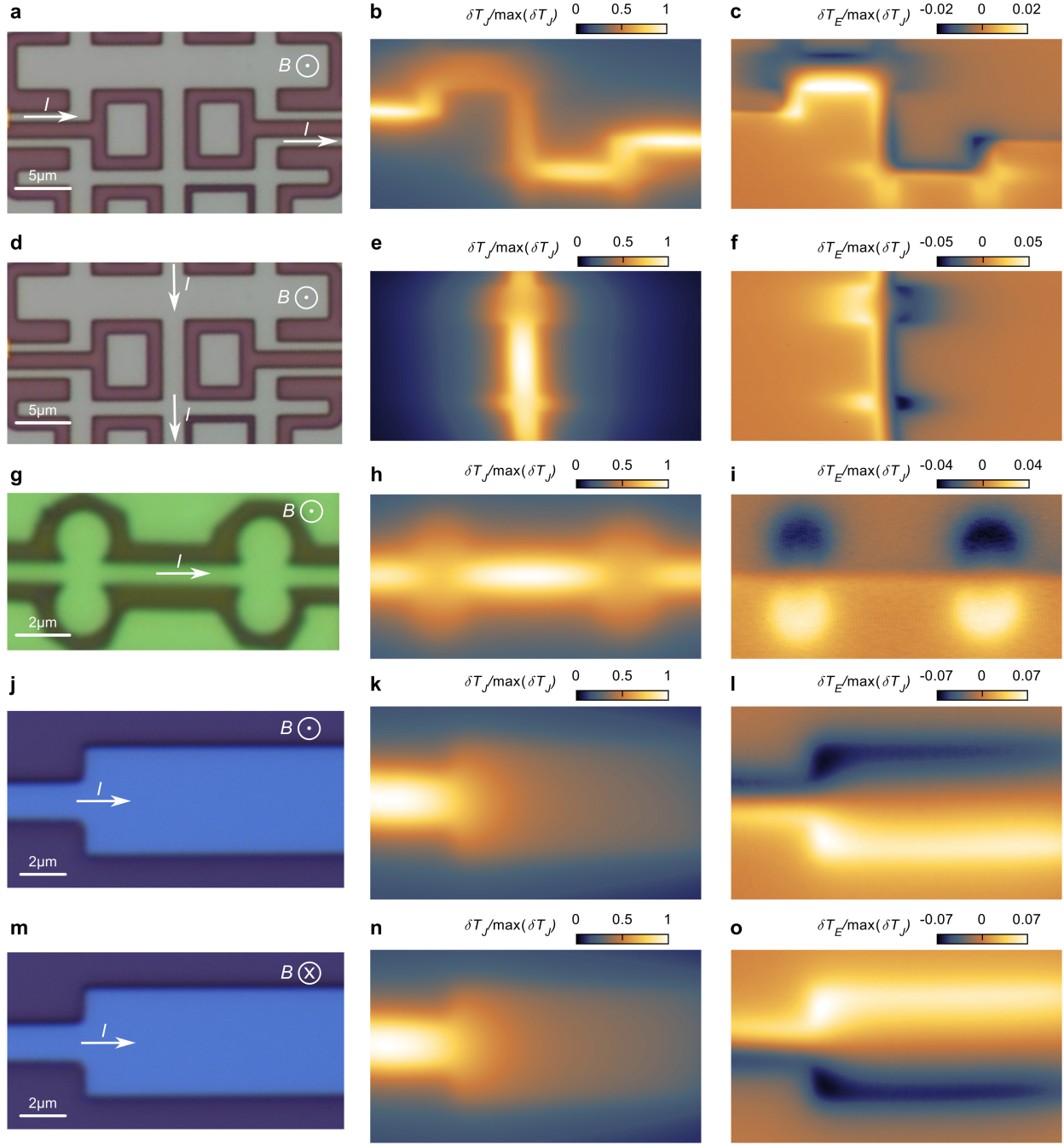

**Extended Data Fig. 4. Thermal imaging in additional geometries.** (**a**) Optical image of a WTe$_2$ sample with strips of different width (grey). The width of the top horizontal strip is $W = 4$ μm and of the lower strip is $W = 1$ μm. The white arrows show the current source and drain. (**b**) Thermal map of $\delta T_J$ normalized by its maximal value $\max(\delta T_J)$ for the current flow configuration in (a). (**c**) The corresponding thermal map of $\delta T_E$ normalized by the same value as in (b). (**d**) Same as a, but with a current configuration along the central vertical strip (white arrows). (**e,f**) Thermal maps of $\delta T_J/\max(\delta T_J)$ and $\delta T_E/\max(\delta T_J)$ respectively for the current configuration in (d). (**g**) Optical image of a WTe$_2$ sample with a narrow strip of width $W = 0.55$ μm connected to circular chambers. (**h,i**) $\delta T_J/\max(\delta T_J)$ and $\delta T_E/\max(\delta T_J)$ for the current configuration in (g). (**j**) Optical image of a WTe$_2$ sample with two regions of different width in series. The magnetic field $\boldsymbol{B}$ points out of the plane. (**k,l**) $\delta T_J/\max(\delta T_J)$ and $\delta T_E/\max(\delta T_J)$ for the configuration in (J). (**m**) Same as (j) but with $\boldsymbol{B}$ pointing in the opposite direction. (**n**) $\delta T_J/\max(\delta T_J)$ for the configuration in (m), showing the same distribution as in (k). (**o**) $\delta T_E/\max(\delta T_J)$ for the configuration in m, showing an opposite temperature gradient relative to (l) due to the inversion of $\boldsymbol{B}$.

22